# BLASTING OPERATIONS USING DIFFERENT INITIATION METHODS IN DEEP UNDERGROUND MINES

Marcin Szumny[1], Piotr Mertuszka[1], Krzysztof Fuławka[1], Eugeniusz Koziarz[2], David Saiang[3]

[1]KGHM CUPRUM Ltd. Research and Development Centre, **Poland**
[2]KGHM Polska Miedź S.A., Rudna Mine, **Poland**
[3]Luleå University of Technology, **Sweden**

**ABSTRACT:**

*Among different types of hazards associated with underground mining in Polish copper mines, one of the most dangerous is rock burst hazard. Having in mind that the depth of exploitation is getting deeper, this problem will likely get worse in the near future. This kind of hazard is connected inherently with seismic events. Controlled group blasting within a potentially unstable roof stratum is considered as an active method of rock burst prevention. A number of recorded seismic events can be clearly and directly explained by the blasting works' effects. With electronic detonators, it is possible to achieve a precise delay time between the detonation of explosives in the individual blastholes and mining faces. Within the framework of this paper, the analysis of seismograms recorded during selected blasting works differing in applied initiation systems, i.e. non-electric and electronic was carried out. It was assumed that this approach may be treated as a new blasting method of rock burst control in deep mines conditions or can be the basis for the modification of the currently used method.*

## 1. INTRODUCTION

An essential aspect of mining is to ensure appropriate working conditions from the safety point of view. The complexity of this problem is particularly evident in the underground mines, where the number of potential threats is high and the effects of dangerous events can be very serious. The cross-section of the risks encountered includes a number of factors, such as ventilation, movement of the machinery and equipment, stability of excavations, etc., the control of which has a significant impact on the safety of works. Due to the increasing depth of mining, geological and mining conditions are becoming more difficult. In the case of Polish copper mines, the depth of mining operations already exceeds 1200 m below the surface. In addition, the increasing mined out area significantly affects the primary stress state within the rock mass. The change in the stress distribution may result in the creation of zones in which the accumulated stress exceed the rock strength, which may lead to dynamic rock failure. This results in a sudden release of elastic energy in the form of tremors or rock bursts. In order to reduce the risk of uncontrolled occurrence of dynamic events a number of technical and organisational measures are generally applied.

Preventive methods can be divided into two groups: (i) passive methods, relate to the organisation or coordination of mining activities and (ii) active methods, which have a direct impact on the state of stresses in the rock mass. Due to the geologic and mining conditions, exploitation of the copper deposits in Poland is carried out using the room-and-pillar mining system with the use of explosives to dislodge the ore (Figure 1).

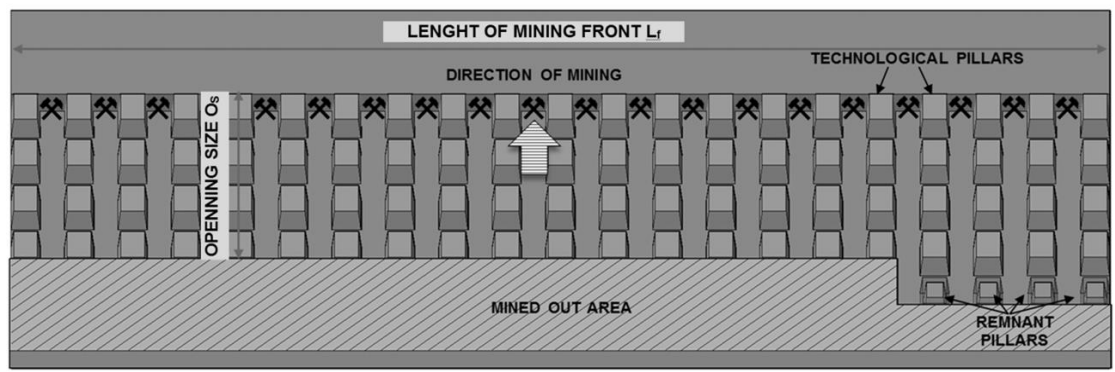

Figure 1. General scheme of the room-and-pillar mining method

One of the most commonly used active methods of rock burst prevention in Polish copper mines is a group blasting of the multiple mine faces. It involves the simultaneous firing of explosives in a large number of faces in order to induce rock mass vibrations [1]. During this time, the crew is outside the danger zone. In this type of blasting, about 15-25 faces are fired on the entire mining front.

Currently, the level of the so called provoked seismic events, which are triggered by blasting works, in relation to spontaneous ones is at the level of 20-30%. So there is still a large potential to increase the effectiveness of this method which may be achieved by means of electronic initiation systems. Due to the accuracy of the electronic detonators, which is significantly higher than electric and non-electric detonators, it has become possible to fire the explosives in a fully controlled manner. The accuracy of electronic initiation systems is around ± 0.01%. The difference in the precision of delays is particularly noticeable compared to non-electric detonators with longer delays. For example, a detonator with a nominal time of 3000 ms and a deviation of ± 2% result in dispersion of absolute delay times of ± 60 ms. In such a case, the probability of the simultaneous firing time of the detonators is very low. As a consequence the holes are fired separately and the actual charge per a given delay significantly decreases. It is worth mentioning that the delay times of non-electric detonators may also differ from the nominal delay times declared by the manufacturer, which was presented in paper [2].

The level of the seismic signal generated by blasting depends mainly on the explosive charge per single delay [3]. Thus, appropriate time synchronisation may have a significant impact on the level of induced vibrations. It may therefore, be assumed that a higher number of simultaneous fired faces will allow for multiplication of this effect, according to the Langefors' relation [4]:

$$V = K \sqrt{\frac{Q}{R^{\frac{3}{2}}}} \qquad (1)$$

where: $V$ – vibration velocity [mm/s]; $R$ – distance [m]; $Q$ – maximum charge per delay [kg]; $K$ – transmission factor.

In view of the above, the use of electronic detonators in appropriate configuration could have a significant impact on paraseismic vibration characteristics [5]. In the case of electronic initiation systems, amplification of vibration amplitude can possibly be achieved in two ways. The first is to increase the amount of explosive that is fired at the same time. In turn, the second is the firing of individual faces in the appropriate time sequence in order to amplify the seismic waves in overstressed areas [6]. This article presents the differences in the characteristics of the seismic signals generated by the face blasting initiated by a non-electric and electronic systems.

2. DESCRIPTION OF TRIAL BLASTING

The drilling and blasting patterns used in destress blasting are in most cases the same as in regular production blasting and must ensure properly formed and fragmented muck pile. Therefore, the basic parameters of this type of blasting are the number of fired faces and the maximum charge per delay. This should, in principle, ensure the amplification of the seismic effect by increasing the amplitude of the induced vibrations. As Mertuszka et al. pointed out, the amplitude of the paraseismic vibrations depends also on the total amount of explosives fired in a given group of mine faces [7]. However, the accuracy of the initiation systems currently used does not provide adequate control of this phenomenon.

The aim of the trials blasting was to verify differences in the vibrations generated by explosive detonation using electronic detonators and modified delay times. Blasting works were carried out at one of the mining panel in Polish copper mine. For comparison purposes, two trials were carried out, in which two different initiation systems and delay times were applied. The aim was to assess the changes in the generated paraseismic waves. Each test consisted of two faces loaded with a bulk emulsion explosive in the amount of about 110 kg per face and on average 3 kg of explosive per hole. The maximum charge per delay in each case was approximately 54 kg. The first two faces were prepared using the typical drilling and blasting pattern with non-electric detonators as shown in Figure 2. Non-electric detonators that are commonly used in the analysed mine were applied for the underground tests.

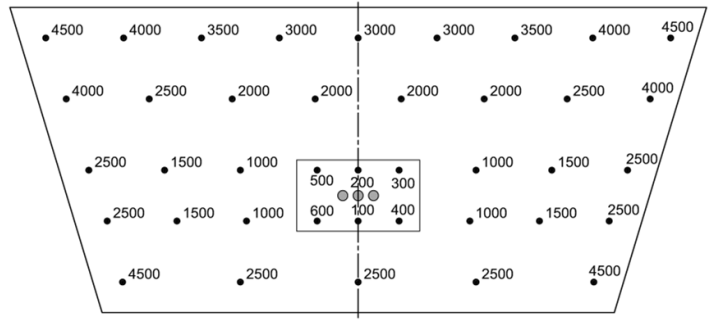

**Figure 2. Drilling and blasting pattern with non-electric detonators (delays in ms)**

The other two faces were initiated using electronic detonators, which were programmed according to the time sequences shown in Figure 3. In each case, the same drilling pattern was used and a burn cut with three empty holes was applied.

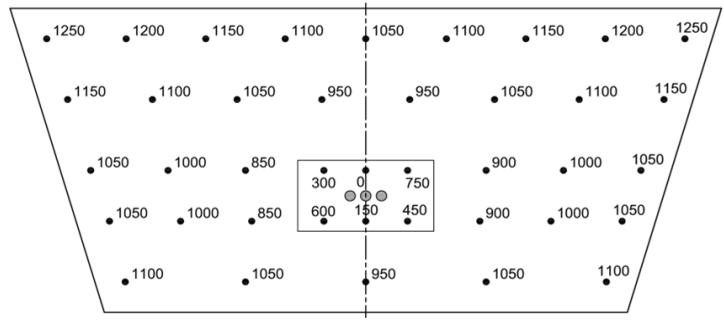

**Figure 3. Drilling and blasting pattern with electronic detonators (delays in ms)**

Vibration velocity measurements were made using uniaxial seismometers (vertical vibration component) with a sampling rate of 500 Hz. The existing seismic network covers the whole mining area. Seismometers were located both on the surface and at the mining level at different distances from the test blasting location. Due to the relatively small amount of explosives, the analysis was based on data recorded at the nearest seismic posts and located around firing site at a distance varying from 700 to 1010 m as shown in Figure 4.

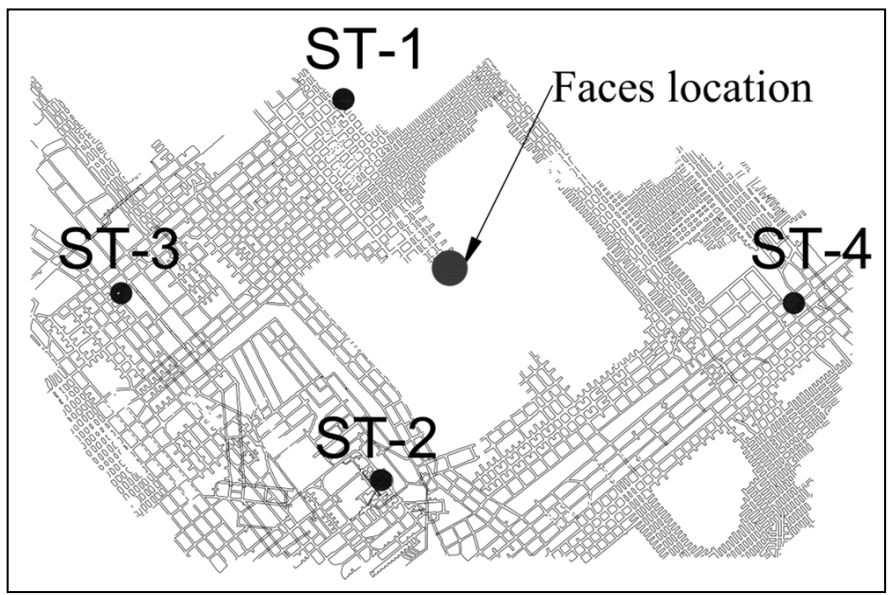

**Figure 4. Location of the seismic posts in relation to the firing site**

In order to avoid the disturbances related to blasting works in other parts of the mine, the trials were carried out separately and independently of regular production blasting. Thanks to this, it was possible to record vibrations from the test firing only.

## 3. ANALYSIS OF RESULTS

For comparative purposes, seismograms recorded at four closest seismometers from ST-1 to ST-4 were used. The analysis was based on a comparison of signals recorded at individual measuring stations and their frequency characteristics. Recorded signals were converted into frequency domain with the use of the Fast Fourier Transform (FFT).

All recorded waveforms for blasting with non-electric detonators and electronic detonators are presented in Figure 5. In addition, values of Peak Particle Velocities (PPV) are shown in table 1.

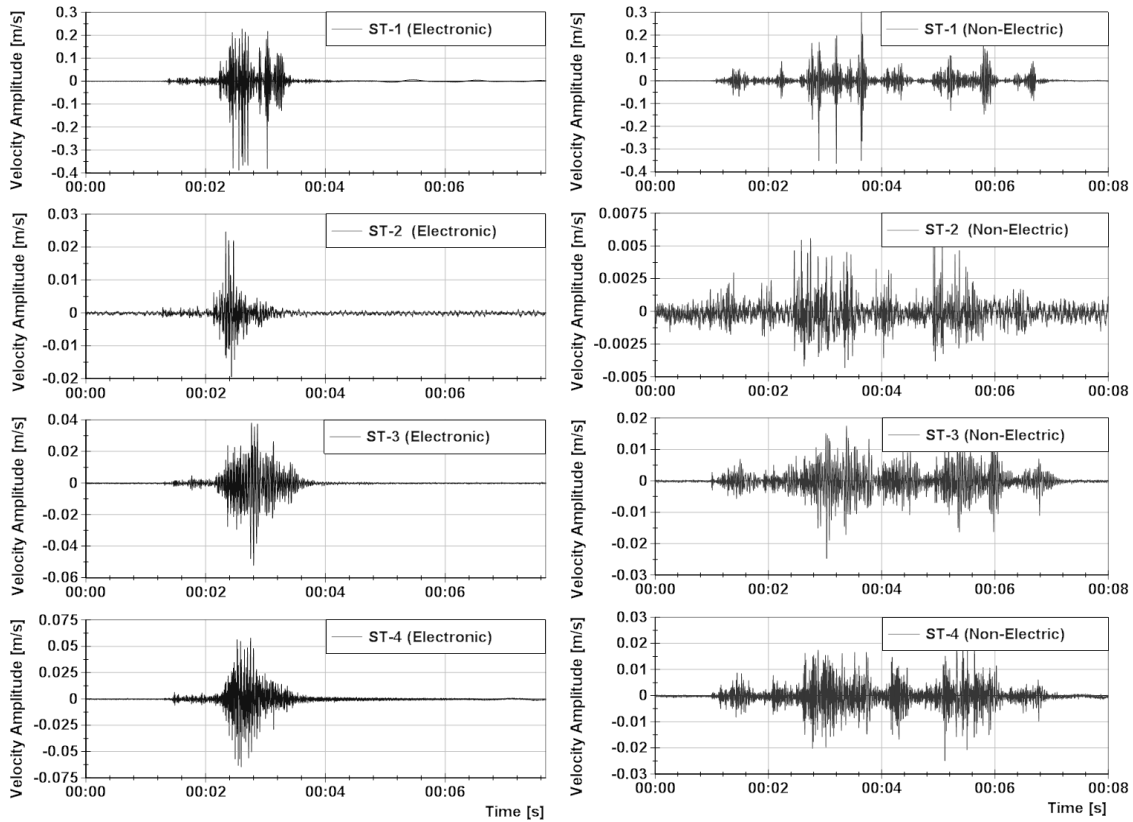

**Figure 5. Vibration velocity waveforms for all seismic posts**

The seismograms clearly show the differences in the time domain characteristic of seismic wave generated by blasting works with different types of detonators. Energy of vibration induced by blasting with using the non-electric detonators dissipates in the longer time in comparison to electronic ones. It significantly affects the effectiveness of active rock burst prevention, where seismic impulse should be as consistent over time as possible. Differences can be also observed when the maximum values of velocity amplitude are considered. In general, seismic effect described by PPV is a few to even dozens of percent higher when electronic detonators are used. The values of recorded PPV are presented in Table 1.

**Table 1. The PPV values and percentage differences for considered initiation methods**

| Seismic post ID | PPV [mm/s] | | Percentage difference [%] |
|---|---|---|---|
| | Electronic | Non-electric | |
| ST-1 | 0.39 | 0.36 | 6.4 |
| ST-2 | 0.02 | 0.01 | 75.4 |
| ST-3 | 0.06 | 0.02 | 61.4 |
| ST-4 | 0.05 | 0.02 | 52.7 |

The maximum vibration amplitudes were recorded at the ST-1 post for each initiation method. In this case, the PPV values were comparable and amounted to 0.36 (non-electric detonators) and 0.39 mm/s (electronic detonators), respectively. Significantly greater differences were observed at other measuring points, where the level of vibrations generated in blasting with electronic detonators were more than 50% higher than for non-electric detonators. The results of spectral analyses indicate the differences in frequency distribution in the recorded signals as shown in Figure 6. Here, a bandpass filter in the range of 1-75 Hz was used.

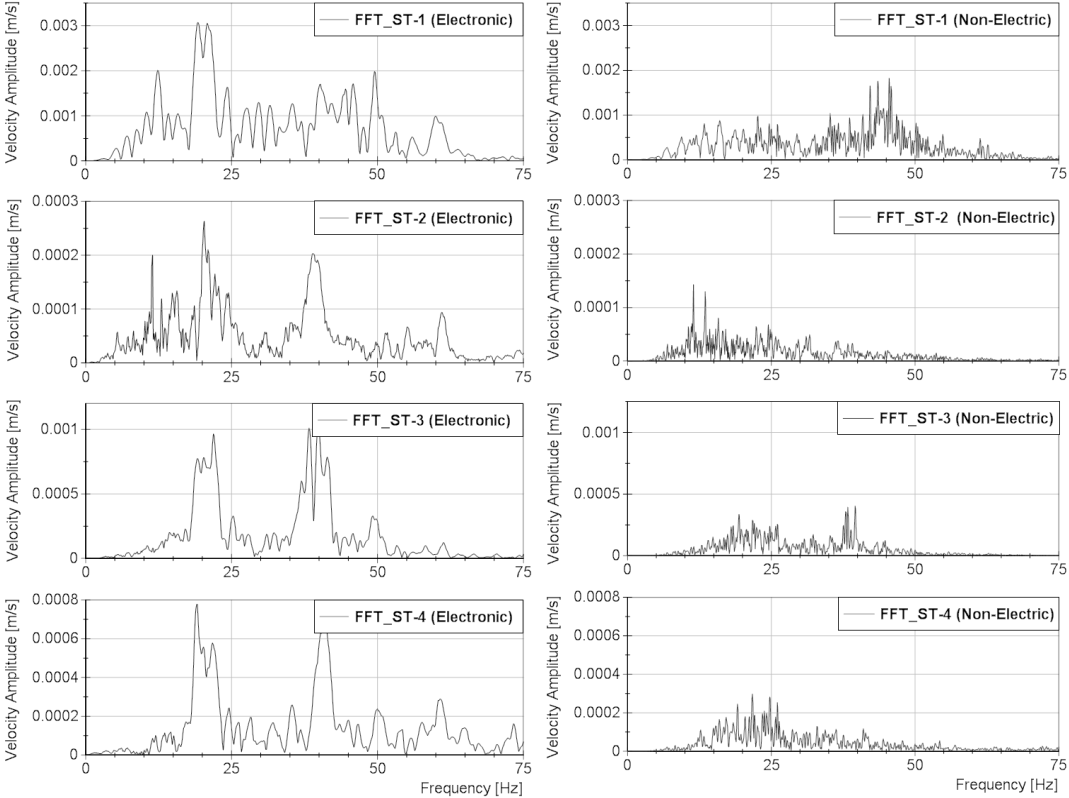

Figure 6. Vibration velocity waveforms for all seismic posts

Based on Figure 6, one may conclude that, the dominant frequency of the signals recorded during blasting with electronic detonators was concentrated in two bands, i.e. 15-25 Hz and 30-40 Hz. In case of initiation with non-electric detonators, the frequency distribution is very diverse and covers a wide frequency band. This clearly shows that the signals generated by blasting using electronic detonators contain clear dominant components, while blasting using non-electric detonators generates signals with components over a wide frequency band without evident dominant components. This can also be seen in the Short-Time Fourier Transform spectrogram shown in Figure 7.

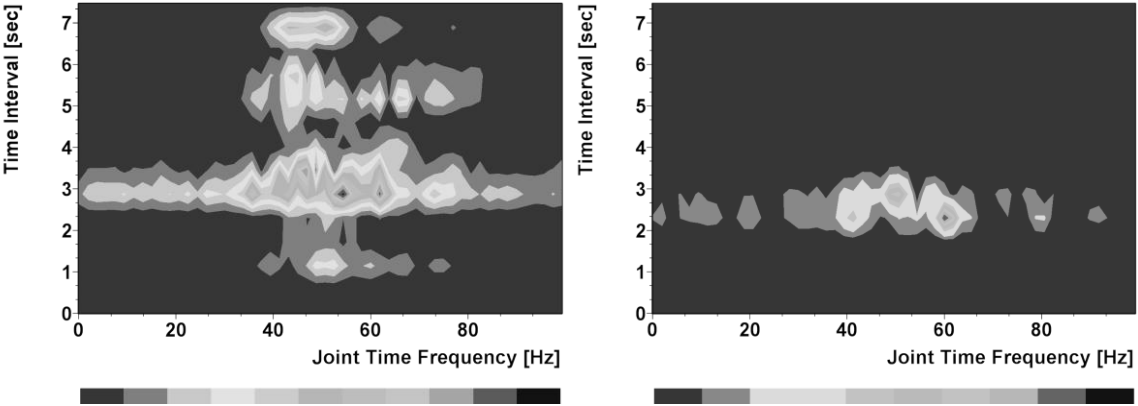

Figure 7. Spectrogram of paraseismic waves (ST-1, non-electric – left, electronic - right)

The frequency band covers almost the entire frequency range. The distribution of signal energy is stretched both in the time and frequency range. The results of STFT analysis of blasting using electronic detonators are presented in the spectrogram in Figure 7.

In blasting using electronic detonators, there is a clear dominant frequency of the signal recorded at the ST-1 station, which is approximately 20 Hz. In addition, the relative power of the signal is concentrated in both time and frequency. The signal is cumulated mainly in the lower frequency range. Therefore, from the provocation point of view, the signal energy is transferred to the rock mass more effectively.

## 4. CONCLUSIONS

The presented analysis proves that electronic initiation systems may be used as an efficient active rock burst prevention method in the condition of Polish copper mines. The structures of generated seismic signals differ significantly depending on the applied initiation system, both in terms of peak amplitudes and frequency characteristics. In the analysed trial with electronic detonators, the PPV values in the most of measurements points were significantly higher than in case of blasting with non-electric detonators. There were also clear differences in frequency characteristic. In case of blasting with electronic detonators, frequency structures were concentrated clearly around two frequencies, namely 20 and 40 Hz and the time of seismic wave was shorter. Blasting with non-electric detonators generated a seismic signal without clear frequency components. What is important, from the mining operation point of view, blast outcomes were similar in both cases.

This could be the basis for modification of destress blasting method in order to release the energy accumulated in the rock mass in the form of tremor or rock burst, and consequently increase the effectiveness in provoking such phenomena. Electronic blasting system could be used as a standard procedure of firing of faces in the group blasting. This may produce the effect of wave amplification due to the interference of post-blasting seismic waves from different mining faces. Due to this fact, the seismic impulse transferred to the rock mass could be much greater than it currently is. All of these aspects open new possibilities for increasing the efficiency of group blasting as an active rock burst prevention method in Polish copper mines. It is worth mentioning that this kind of development is relatively simple for implementation in Polish copper mines.

## ACKNOWLEDGEMENTS

**This paper has been prepared through the Horizon 2020 EU funded project on "Sustainable Intelligent Mining Systems (SIMS)", Grant Agreement No. 730302.**